# Thermodynamic phase stability, structural, mechanical, optoelectronic, and thermoelectric properties of the III-V semiconductor AlSb for energy conversion applications

Iskandar Raufov [a], Dilshod Nematov [a,b,*], Saidjafar Murodzoda [a], Sakhidod Sattorzoda [a], Anushervon Ashurov [a]

[a] *S.U. Umarov Physical-Technical Institute of NAST, Dushanbe 734063, Tajikistan*
[b] *School of Optoelectronic Engineering & CQUPT-BUL Innovation Institute, Chongqing University of Posts and Telecommunications, Chongqing 400065, China*

ABSTRACT

This study presents a first-principles investigation of the structural, thermodynamic, electronic, optical, and thermoelectric properties of aluminum antimonide (AlSb) in its cubic (F-43 m) and hexagonal (P6$_3$mc) phases. Both structures are dynamically and mechanically stable, which is confirmed by phonon calculations and the Born-Huang criteria. The lattice constants obtained using the SCAN and PBEsol functionals show good agreement with experimental data. The cubic phase has a direct band gap of 1.66–1.78 eV, while the hexagonal phase has 1.48–1.59 eV, as confirmed by mBJ and HSE06 calculations. Under external pressure, the band gap decreases in the cubic phase and increases in the hexagonal phase due to different s-p orbital hybridization mechanisms. The optical absorption coefficient reaches $10^6$ cm$^{-1}$, which is comparable to or higher than values known for other III–V semiconductors. The Seebeck coefficient exceeds 1500 µV/K under intrinsic conditions, and the thermoelectric performance improves above 600 K due to enhanced phonon scattering and lattice anharmonicity. The calculated formation energies (-1.316 eV for F-43 m and -1.258 eV for P6$_3$mc) confirm that the cubic phase is thermodynamically more stable. The hexagonal phase shows higher anisotropy and lower lattice stiffness, which is beneficial for thermoelectric applications. These results demonstrate how crystal symmetry, phonon behavior, and charge transport are connected, and they provide useful guidance for the design of AlSb-based materials for optoelectronic and energy-conversion technologies.

## 1. Introduction

The development of modern technologies for energy conversion and information processing critically depends on the search for novel semiconducting materials with a comprehensive set of functional properties. In recent decades, halide perovskites, kesterites, CdTe, Si, GaAs, and related compounds have been extensively studied, achieving remarkable efficiencies in photovoltaic, thermoelectric, and optoelectronic devices [1–4]. However, despite their success, many of these materials face serious limitations, including poor thermal stability, moisture and UV degradation, phase instability, toxicity of certain components, and high production costs [5–7]. These limitations motivate the search for alternative semiconductors with improved reliability, technological compatibility, and large-scale applicability.

Aluminum antimonide (AlSb), a III-V semiconductor historically known for its optoelectronic performance, has recently regained attention due to its unique combination of characteristics. It possesses a relatively wide band gap, high carrier mobility, and compatibility with III-V heteroepitaxy, making it attractive for infrared optoelectronics, photovoltaics, and thermoelectrics [8–10]. Phase engineering represents a particularly important aspect. While the cubic zinc-blende (F-43 m) phase is thermodynamically stable, the hexagonal wurtzite-like (P6$_3$mc) phase can be stabilized under epitaxial growth or nanoscale confinement, offering the potential for band gap tuning and modified functionalities [11–13]. Thermoelectric properties of bulk AlSb have traditionally been limited by relatively high lattice thermal conductivity and suboptimal conductivity/Seebeck balance, resulting in low ZT values [14]. Nevertheless, recent advances such as multiphonon scattering control, doping, composites, and nanostructuring provide pathways for improved performance [15–17]. Parallel studies on structurally related Zintl phases (for example NaSrSb and NaBaSb) have reported ZT ≈ 2.0 at 900 K, attributed to ultralow lattice thermal conductivity and high Seebeck coefficients [18]. Similar design strategies have also been successfully applied to complex double perovskites ($A_2AuScX_6$), Ca$(InP)_2$, and doped HgSe alloys, highlighting the broader relevance of structural-electronic tuning [19–23]. Recent studies have also reported significant progress in understanding the structural, optoelectronic, and thermoelectric behavior of AlSb and related III-V semiconductors. For example, pressure-driven band-gap engineering, defect-tolerant optical transitions, and mechanical tunability were demonstrated in several recent works [20–24]. These studies highlight the strong sensitivity of III-V compounds to bonding asymmetry, local coordination, and phase-dependent orbital hybridization, and emphasize the need for systematic comparative analyses of cubic and hexagonal polymorphs.

Recent studies of two-dimensional AlSb monolayers confirm a direct band gap of 0.93–1.65 eV, high carrier mobility, as well as strong excitonic effects and significant many-body interactions shaping the optical response [24]. For bulk crystals, pressure-induced structural transitions into denser NaCl-type (B1) and CsCl-type (B2) phases have been reported, accompanied by changes in electronic structure and mechanical properties [25–27]. Yet, systematic studies on phase competition between cubic (F-43 m) and hexagonal (P6$_3$mc) modifications remain scarce, with little data on relative total energies, phonon stability, thermodynamic properties, and complete sets of elastic constants [28].

In terms of practical opportunities, AlSb combines several advantages: technological compatibility with III-V devices, potential for infrared photodetectors and optoelectronics, tunable band structure via strain and doping, and radiation resistance, which is especially relevant for aerospace and defense applications [29]. In thermoelectrics, thermal transport processes play a decisive role. Recent works demonstrated that four-phonon scattering can strongly reduce the predicted thermal conductivity of III-V compounds (including AlSb), reshaping the understanding of intrinsic transport limits and offering new strategies for phonon engineering [15–17].

Despite this progress, several open questions remain, as comprehensive data are still lacking for the phonon stability of the hexagonal phase, its mechanical and thermoelectric characteristics, and the influence of defects and impurities on electronic and optical properties [28, 29]. Thermal transport processes, particularly multiphonon scattering, remain insufficiently explored, although they play a crucial role in determining the thermoelectric figure of merit [5,30]. These gaps highlight the need for further investigations to uncover the hidden potential of AlSb as a multifunctional semiconductor of the next generation.

In this work, we perform a comprehensive first-principles study of the structural, mechanical, electronic, optical, thermodynamic, and thermoelectric properties of AlSb in both F-43 m and P6$_3$mc phases, with emphasis on pressure-dependent stability and multifunctional potential.

## 2. Computational details

In this study, the structural, electronic, optical, mechanical, thermodynamic, and thermoelectric properties of aluminum antimonide (AlSb) in the cubic F-43 m and hexagonal P6$_3$mc phases were investigated using density functional theory (DFT) as implemented in the VASP package [31]. The interactions between electrons and ions were described using the projector augmented-wave (PAW) method [32], with explicit consideration of the valence electron configurations: Al: $3s^23p^1$ and Sb: $4d^{10}5s^25p^3$. Structural optimization was performed using the SCAN functional [33] until the residual forces were below 0.01 eV/Å and the total energy convergence threshold was set to $10^{-6}$ eV. Convergence was achieved with a plane-wave cutoff energy of 700 eV and a Γ-centered k-point mesh of 7 × 7 7 × 7 for the cubic F-43 m phase, and 12 12 × × 12 12 × 5 for the hexagonal P6$_3$mc phase of AlSb. To obtain a reliable estimation of the electronic band gap, both the modified Becke-Johnson (mBJ) potential [34] and the hybrid HSE06 functional [35] were employed. The total and partial densities of states (DOS and PDOS), electronic band structures, and optical properties were calculated using the mBJ potential based on the SCAN-relaxed structures. Optical characteristics, including the dielectric function, absorption coefficient, and refractive index, were determined within the linear response formalism. Mechanical properties were evaluated using the GGA-PBE functional while preserving the SCAN-optimized geometry. The second-order elastic constants were obtained through the finite-strain method, and mechanical stability was assessed according to the Born-Huang criteria for cubic and hexagonal crystals. Phonon and thermodynamic properties were analyzed using the Phonopy package interfaced with VASP 6.4.3 via the finite-displacement method in a 2 × 2 × 2 supercell. From the phonon density of states, Helmholtz free energy, entropy, and constant-volume heat capacity were calculated in the temperature range of 0–1000 K. The absence of imaginary frequencies in the phonon spectra confirmed the dynamical stability of all examined phases. Thermoelectric properties were evaluated using the BoltzTraP2 code interfaced with Quantum ESPRESSO 7.4.1 [36].

## 3. Results and discussion

### 3.1. Structural properties

Aluminum antimonide crystallizes in two structural modifications: a cubic phase with space group F-43 m and a hexagonal phase with space group P6$_3$mc. To ensure the reliability and accuracy of subsequent calculations, convergence tests were first performed to determine the optimal values of the plane-wave cutoff energy (ENCUT) and the k-point sampling density for Brillouin zone integration. Convergence with respect to both parameters was analyzed using the Monkhorst-Pack and Γ-centered schemes, allowing the selection of appropriate computational settings. As shown in Figure S1, once the cutoff energy reached 700 eV, the total energy of the system became stable for both phases under either sampling scheme. Further increases in ENCUT produced negligible changes in total energy while significantly increasing computational cost. This convergence approach is robust: despite minor variations in total energy behavior with respect to the k-point density, both sampling methods exhibit excellent convergence characteristics. This confirms the correctness of the chosen Brillouin zone discretization parameters and the reliability of the obtained computational results. In both phases, cubic and hexagonal, the Monkhorst-Pack and Γ-centered schemes yield nearly identical total energies, indicating the consistency and stability of the adopted electronic-structure modeling methodology.

Similarly, for both AlSb phases, k-point optimization was performed at ENCUT = 315 eV (approximately 1.3 × ENMAX), where the optimal meshes were determined to be 7 × 7 × 7 (Γ-centered) for the cubic F-43 m phase and 12 × 12 × 5 (Γ-centered) for the hexagonal P6$_3$mc phase (Figure S2). These optimized ENCUT and k-point parameters were subsequently used as the standard computational setup for electronic-structure and optical-property calculations. Based on our observations, the Monkhorst-Pack grid ensures uniform Brillouin zone sampling without necessarily including the Γ point, which can be efficient for highly symmetric periodic systems. In contrast, the Γ-centered mesh explicitly includes the Γ point and often provides superior convergence for systems with lower symmetry or when accurate description of states near the Γ point is essential.

In the performed calculations, both discretization schemes exhibited excellent convergence of the total energy. The difference in total energy per atom between the two methods was <1 meV, confirming that the chosen k-point density ensures sufficient numerical accuracy without unnecessary computational expense. After establishing the optimal computational parameters, full structural optimizations were carried out for both AlSb phases. The equilibrium lattice constants, unit cell volumes, and interaxial angles were determined and compared across different exchange-correlation functionals as well as with available experimental data (Table 1). The PBE functional belongs to the generalized gradient approximation (GGA) family, while PBEsol and SCAN are also GGA-based, with SCAN representing a more advanced meta-GGA formulation that improves accuracy for both structural and energetic predictions.

**Table 1**
Lattice parameters of cubic and hexagonal AlSb phases obtained with various exchange-correlation functionals.

| AlSb | Lattice constants | PBE (GGA) | PBEsol | LDA | SCAN | Experimental |
|---|---|---|---|---|---|---|
| F43m | a (Å) | 6.233 | 6.169 | 6.121 | 6.17 | 6.14 [37,38] |
|  | α=β=γ, ° | 90.00 | 90.00 | 90.00 | 90.00 |  |
|  | V (Å³) | 242.15 | 234.77 | 229.33 | 235.11 |  |
| P6₃mc | a (Å) | 4.39 | 4.34 | 4.31 | 4.35 |  |
|  | b (Å) | 3.80 | 3.76 | 3.73 | 3.76 |  |
|  | c (Å) | 7.23 | 7.16 | 7.10 | 7.17 |  |
|  | α= β≠γ, ° | 90, 12 | 90, 12 | 90, 12 | 90, 12 |  |
|  | V (Å³) | 120.99 | 117.30 | 114.57 | 117.55 |  |

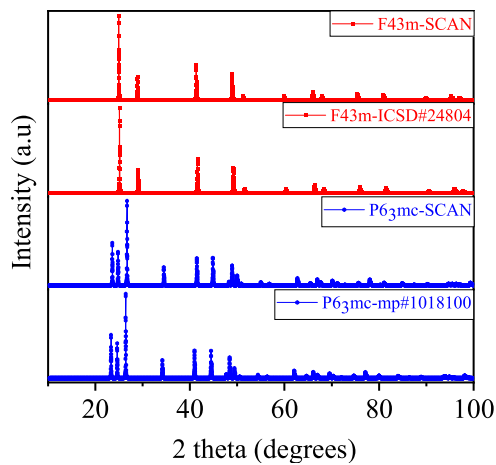

**Fig. 1.** Calculated X-ray diffraction (XRD) patterns for the cubic (F-43 m) and hexagonal (P6₃mc) phases of AlSb, obtained using Cu-Kα radiation and visualized with the *VESTA* software.

A comparative analysis of the lattice parameters reveals a consistent trend across both phases of AlSb: the PBE functional predicts the largest lattice constants, while LDA gives the smallest. The SCAN-derived lattice constant for the cubic phase ($a$ = 6.17 Å) matches closely the experimental values of 6.14–6.15 Å, confirming the accuracy of the structural optimization [37,38]. The hexagonal phase also yields realistic lattice parameters that fall within the expected range predicted for wurtzite-like III–V structures, consistent with reported theoretical studies. The values obtained using PBEsol and SCAN lie between these two limits and are in closer agreement with experimental data. For example, for the cubic phase of AlSb, the calculated lattice constants are $a$ = 6.233 Å (PBE), $a$ = 6.169 Å (PBEsol), $a$ = 6.121 Å (LDA), and $a$ = 6.172 Å (SCAN). A similar trend is observed for the hexagonal phase. These differences arise from the intrinsic characteristics of the respective exchange-correlation functionals. PBE is known to slightly overestimate lattice volumes due to its weaker treatment of correlation effects. In contrast, LDA assumes a homogeneous electron gas approximation, which often results in underestimated lattice parameters. PBEsol, specifically designed for solid-state systems, provides more accurate equilibrium volumes, whereas SCAN, as a meta-GGA functional, offers improved reliability in predicting both structural and electronic properties.

Fig. 1 shows the calculated theoretical X-ray diffraction (XRD) patterns for both AlSb phases, simulated using Cu-Kα radiation within the VESTA software package. The well-defined diffraction peaks confirm the high crystallinity of both structures and are consistent with the expected symmetry characteristics of the cubic F-43 m and hexagonal P6₃mc phases.

The simulated diffraction profiles exhibit the characteristic peaks corresponding to the F-43 m and P6₃mc space groups, confirming the crystalline purity and the absence of secondary phases. The cubic phase, shown in red, was compared with the ICSD database, revealing a complete match of all major reflections without the appearance of additional peaks, which verifies the phase purity of the calculated cubic AlSb. The hexagonal phase, plotted in blue, was validated against the Materials Project database due to the absence of relevant ICSD data. The lack of extra peaks likewise confirms the structural purity of the hexagonal AlSb phase.

### 3.2. Phonon and thermodynamic properties

To assess the stability and thermal behavior of the AlSb phases, phonon, thermodynamic, and elastic properties were calculated. These parameters provide insights not only into the dynamical stability of the crystal lattice but also into its heat transport behavior and vibrational anharmonicity. Figures 2(a-d) present the phonon dispersion relations and phonon density of states (PhDOS) for the cubic and hexagonal phases of AlSb, calculated along the high-symmetry directions of the Brillouin zone. The absence of imaginary frequencies in both dispersion spectra (Figs. 2a,b) indicates that the optimized structures are dynamically stable, confirming the lack of soft modes that could lead to lattice instability. In the cubic phase, the optical phonon branches extend up to approximately 10 THz, suggesting stiffer interatomic bonding and a more rigid lattice. By contrast, the phonon spectrum of the hexagonal phase is shifted toward lower frequencies, reflecting softer lattice dynamics and higher vibrational anisotropy. In the phonon density of states, the low-frequency region (2–4 THz) is primarily dominated by vibrations of the heavier Sb atoms, while the high-frequency region (8–10 THz) corresponds to the lighter Al atoms (Figs. 2c,d). This mass-dependent separation of vibrational modes is typical for binary compounds with significant atomic mass contrast and plays a crucial role in defining their thermal transport characteristics.

Both AlSb phases are phonon-stable, yet the hexagonal structure exhibits lower stiffness, which can be advantageous for thermoelectric applications due to its potentially reduced lattice thermal conductivity. The sharper peaks observed in the Phonon DOS of the cubic phase imply a higher degree of atomic ordering and stronger covalent bonding, whereas the broader, smoother spectra of the hexagonal phase suggest enhanced lattice anharmonicity, which promotes phonon-phonon scattering and consequently lowers thermal conductivity.

Thus, both phases of AlSb exhibit no imaginary frequencies, confirming their dynamical stability in agreement with the thermodynamic results. The main phonon characteristics of both phases, including the range of acoustic modes and maximum vibrational frequencies, are summarized in Table 2.

The obtained data indicate that the cubic phase is characterized by higher phonon frequencies, reflecting stronger interatomic bonding and higher lattice thermal conductivity. In contrast, the hexagonal modification exhibits a softer lattice and pronounced vibrational anisotropy, which can reduce heat transport and enhance thermoelectric performance.

Subsequently, the temperature-dependent thermodynamic functions were analyzed in the range of 0–1000 K, including entropy S(T), Helmholtz free energy F(T), and constant-volume heat capacity Cv(T). These quantities were calculated from the phonon density of states g(ω) within the harmonic approximation, according to the following expressions:

$$F(T) = \int_0^\infty g(\omega)\left(\frac{1}{2}\hbar\omega + k_B T \ln\left(1 - e^{-\hbar\omega/k_B T}\right)\right)d\omega,$$

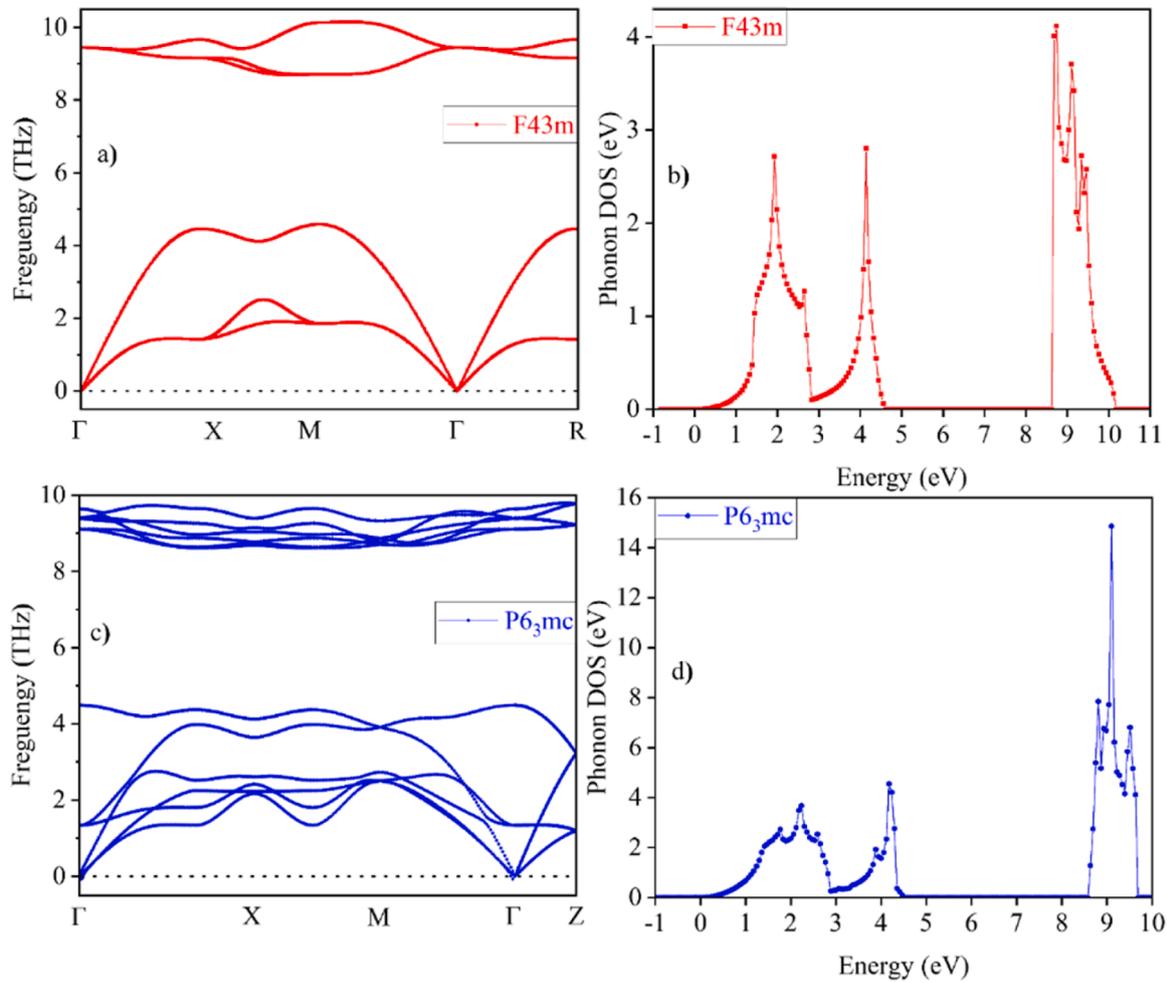

**Fig. 2.** Calculated (a) phonon dispersion curves and (b) phonon density of states (Phonon DOS) for the cubic (F-43 m) and hexagonal (P6$_3$mc) phases of AlSb.

**Table 2**
Phonon characteristics of the AlSb phases.

| AlSb | Maximum Phonon Frequency (THz) | Acoustic Branch Range (THz) | tability Indicator (Imaginary Frequencies) | Key Feature |
|---|---|---|---|---|
| F-43m | ≈ 10 | 0–4 | Absent | Rigid lattice with high structural stability |
| P6$_3$mc | ≈ 8.5 | 0–3.5 | Absent | Soft lattice with pronounced anisotropy |

**Table 3**
Comparison of thermodynamic characteristics of AlSb at 300, 600, and 900 K.

| T, (K) | Cv (J·mol$^{-1}$·K$^{-1}$) | S (J·mol$^{-1}$·K$^{-1}$) | F (eV/ formula unit) | H (eV/ formula unit) | Remark |
|---|---|---|---|---|---|
| 300 | 23–24 | 60–62 | −18.92 | −18.84 | Phonon contribution indicates lattice stabilization |
| 600 | 28–29 | 95–100 | −18.70 | −18.50 | Cv approaches the Dulong-Petit limit |
| 900 | 30–31 | 130–135 | −18.45 | −18.10 | Anharmonicity increases; thermal stability maintained |

$$S(T) = -\int_0^\infty g(\omega)[n(\omega,T)\ln n(\omega,T) - (1+n(\omega,T))\ln(1+n(\omega,T))]\,d\omega,$$

$$C_v(T) = k_B \int_0^\infty g(\omega)\left(\frac{\hbar\omega}{k_B T}\right)^2 \frac{e^{\hbar\omega/k_B T}}{(e^{\hbar\omega/k_B T} - 1)^2}\,d\omega,$$

where $n(\omega,T) = \frac{1}{e^{\hbar\omega/k_B T} - 1}$ - is the Bose-Einstein distribution, $\hbar$ is the reduced Planck constant, $k_B$-is the Boltzmann constant and $\omega$ represents the phonon frequency.

The calculations show that the thermodynamic behavior of both AlSb phases follows the classical anharmonic models and approaches the Dulong-Petit limit at high temperatures. This trend reflects the expected saturation of the heat capacity and the stabilization of vibrational entropy with increasing temperature. The generalized numerical values of the thermodynamic parameters at selected characteristic temperatures are summarized in Table 3.

In Fig. 3a, which illustrates the temperature dependence of entropy, a monotonic increase in S(T) is observed for both phases, as expected. This behavior is attributed to the progressive activation of phonon modes with increasing temperature and the corresponding rise in lattice disorder.

The absolute entropy values show moderate differences between the two phases, which can be explained by variations in their structural rigidity and vibrational characteristics. The sensitivity of entropy to crystallochemical parameters thus makes it a valuable indicator for evaluating energy distribution and lattice stability under thermal

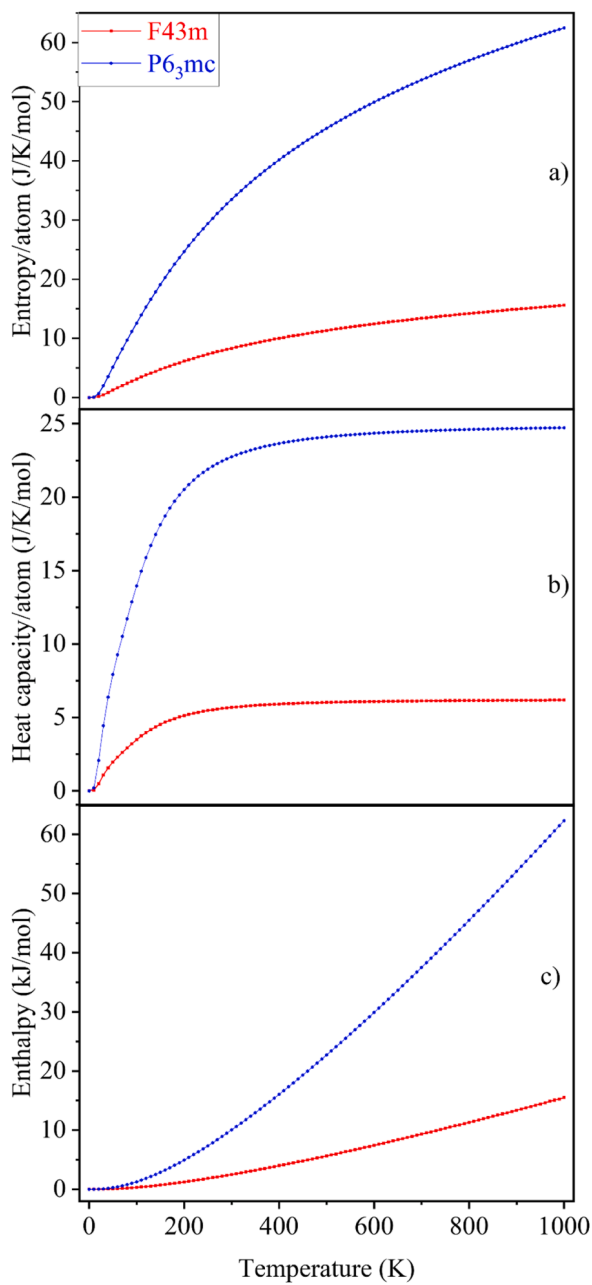

**Fig. 3.** Temperature dependence of (a) entropy, (b) heat capacity at constant volume, and (c) enthalpy for the cubic and hexagonal phases of AlSb.

excitation.

In Fig. 3b, the variation of constant-volume heat capacity with temperature shows that both phases exhibit the characteristic behavior predicted by quantum solid-state theory: a rapid increase in heat capacity at low temperatures followed by saturation at higher temperatures approaching the classical limit. The differences between the two curves reflect variations in the low-frequency phonon density of states, which predominantly governs heat capacity in the low-temperature region. This comparison highlights how the structural features of the crystal lattice directly influence the material's ability to store thermal energy, a topic further discussed in the section on the thermoelectric properties of AlSb. Fig. 3c illustrates the temperature dependence of enthalpy, which increases steadily with temperature. This growth indicates that, upon heating, the system absorbs a certain amount of thermal energy, leading to an increase in the internal energy of atoms and phonons. As the temperature rises, atomic vibrations intensify, producing a pronounced increase in enthalpy. The continuous rise of H(T) with temperature confirms that AlSb remains thermodynamically stable throughout the investigated range.

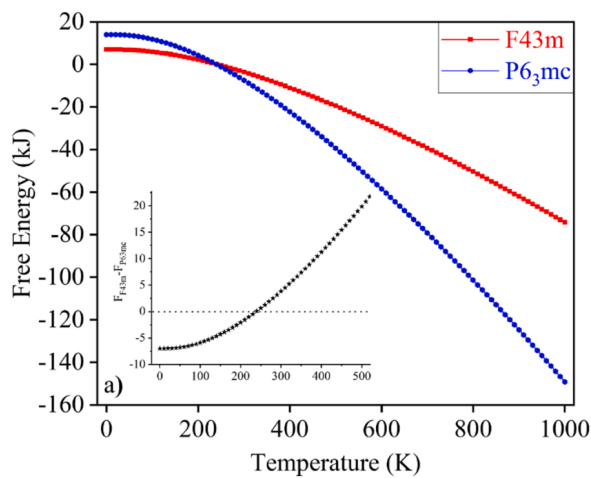

**Fig. 4.** Temperature dependence of the Helmholtz free energy of AlSb.

Fig. 4 presents the Helmholtz free energy F(T) as a function of temperature, representing the total thermodynamic potential of the system. The observed decrease in F(T) with increasing temperature for both phases results from the growing entropic contribution and aligns with the general thermodynamic behavior expected for condensed matter systems.

Although the differences between the two phases are relatively small, they persist and become more pronounced at higher temperatures, indicating distinct degrees of thermodynamic stability. As shown in Fig. 4, AlSb exhibits a phase transition near 245–250 K. Lower values of the Helmholtz free energy correspond to a more stable configuration at a given temperature, which is particularly important when evaluating materials for high-temperature applications. The analysis confirms the overall stability of both phases across a wide temperature range. Additionally, the calculated formation energies are $-1.316$ eV for the F-43m and $-1.258$ eV for the P6$_3$mc phase, confirming the higher thermodynamic stability of the cubic modification. Direct comparison with literature values is difficult because reliable experimental formation energies for AlSb are not available, and existing theoretical reports for the cubic phase vary considerably depending on the exchange–correlation functional and reference potentials. These results are consistent with the phonon and thermodynamic analyses and establish a foundation for future thermodynamic modeling, including phase diagram construction and assessment of stability during the formation of solid phases and heterostructures. For a more detailed understanding of the thermodynamic and phonon-related stability, Figs. 9a-b present the calculated pressure-dependent elastic moduli ($C_{11}$, $C_{12}$, $C_{44}$, etc.) within the 0–30 GPa range for both phases. The complete set of elastic constants is provided in Table S2.

### 3.3. Electronic structure of alsb

After comprehensive structural optimization, the electronic band structures of both AlSb phases were computed and analyzed to evaluate key electronic parameters such as band gap, band dispersion, and total and partial density of states. Based on the well-optimized geometries obtained using GGA, PBEsol, LDA, and SCAN functionals, the band gaps were further refined employing the hybrid HSE06 functional and the modified Becke-Johnson (mBJ) potential. The comparative results are presented in Table 4.

The obtained results show that, regardless of the exchange-correlation functional used, both the hybrid HSE06 and the modified Becke-Johnson (mBJ) potential yield band gap values that are in close agreement with each other and consistent with experimental observations. Spin–orbit coupling (SOC) was not included because previous studies showed that SOC induces only a small shift in the band edges of AlSb due to the relatively light mass of Al and moderate SOC strength of Sb. Therefore, the direct bandgap character and optical transitions remain unchanged, and the mBJ/HSE06 results remain reliable [42]. For the cubic F-43 m phase, the calculated band gap varies between 1.66 and 1.78 eV, which agrees well with the experimental range of 1.63–1.81 eV [39–41]. This consistency confirms the reliability of the selected computational methods and the accuracy of the treatment of exchange-correlation effects in the AlSb system.

Table 4
Comparison of the calculated band gap (Eg) values for the cubic and hexagonal phases of AlSb using the HSE06 and mBJ methods.

| AlSb | HSE06/LDA | HSE06/PBE | HSE06/PBEsol | HSE06/SCAN | mBJ/LDA | mBJ/PBE | mBJ/PBEsol | mBJ/SCAN | EXPERIMENTAL |
|---|---|---|---|---|---|---|---|---|---|
| F43m | 1.715 | 1.781 | 1.737 | 1.748 | 1.661 | 1.716 | 1.672 | 1.683 | 1.63[39], 1.75[40], 1.81[41] |
| P6$_3$mc | 1.595 | 1.496 | 1.584 | 1.584 | 1.529 | 1.408 | 1.485 | 1.485 | - |

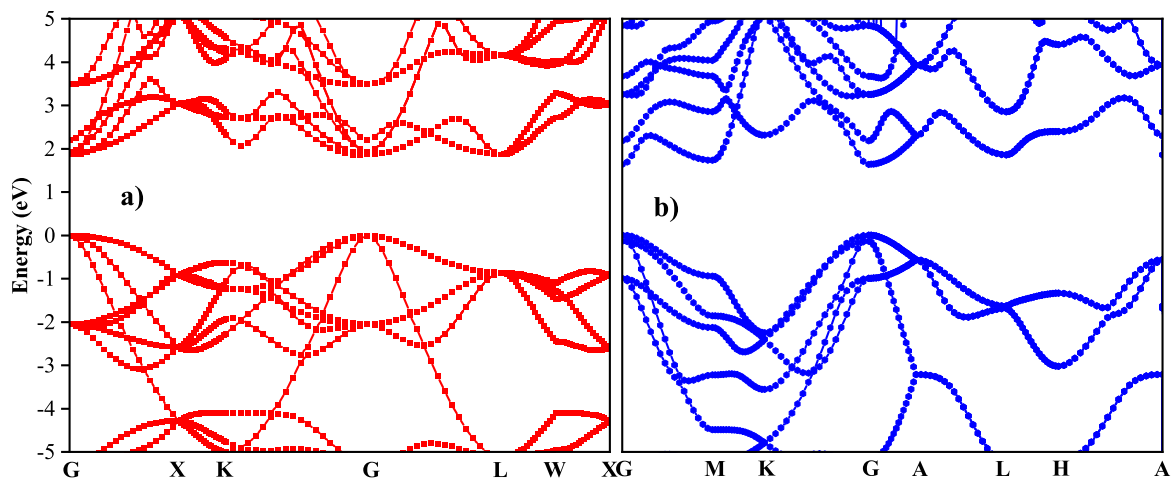

**Fig. 5.** Calculated electronic band structures of cubic (a) and hexagonal (b) phases of AlSb.

In contrast, the hexagonal P6$_3$mc modification exhibits a slightly narrower band gap, indicating stronger overlap between the valence and conduction bands and potentially higher carrier mobility. The reduction of Eg compared with the cubic phase arises from changes in local coordination and reduced lattice symmetry, which result in denser packing of electronic states and weaker hybridization between the Al and Sb s and p orbitals. Comparison across different functionals shows that LDA and PBE systematically underestimate Eg due to self-interaction errors, while SCAN and particularly HSE06 provide more realistic results. The mBJ potential demonstrates comparable accuracy to HSE06, differing by only 0.05–0.07 eV, but with significantly lower computational cost. These findings confirm that both HSE06 and mBJ are reliable tools for accurately describing the electronic band structure of AlSb and analyzing how crystallochemical factors affect the band gap in both phases. The strong agreement between HSE06 and mBJ has been repeatedly validated in previous studies [42–59].

The band structure diagrams presented in Fig. 5, plotted along the high-symmetry directions of the Brillouin zone, reveal that both AlSb phases possess a direct band gap transition from the valence to the conduction band, which is a crucial property for modern optoelectronic devices. The analysis shows that both phases behave as semiconductors with a well-defined separation between valence and conduction states. The cubic phase exhibits a slightly wider band gap, whereas the hexagonal phase has a narrower one, consistent with the results in Table 4. Both structures feature an extended electronic dispersion, an important factor for improving the performance of photovoltaic materials since the main part of the solar spectrum falls within this energetic region.

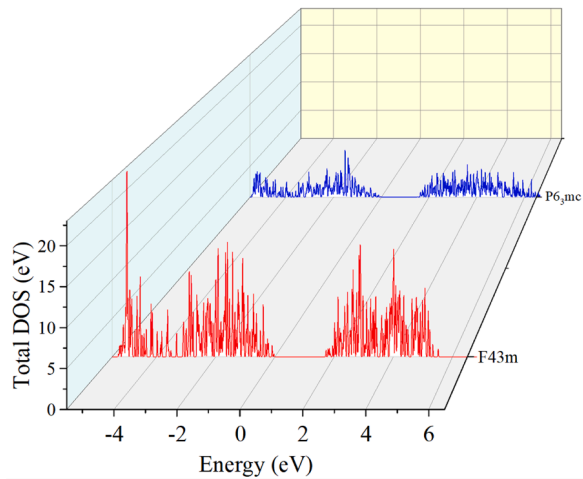

**Fig. 6.** Calculated total electronic density of states (DOS) of AlSb using the HSE06 functional.

The presence of a direct band gap promotes efficient generation and recombination of electron-hole pairs in the studied material. This property enhances both light absorption and photoluminescence, which are critical for the performance of optoelectronic devices such as light-emitting diodes (LEDs) and lasers. Moreover, materials with a wide and direct band gap can reduce energy losses caused by thermal conductivity and enable stable operation of electronic components under high voltage, an important aspect in the design of photovoltaic modules.

Furthermore, the calculated density of states (DOS) complements the band structure diagrams shown in Fig. 5 and confirms the observed electronic trends. The DOS analysis provides additional insight into the evolution of electronic properties, offering a consistent description of the orbital contributions and hybridization effects in both phases (Fig. 6).

As shown in Fig. 6, the transition from the cubic to the hexagonal phase leads to a narrowing of the band gap. The primary distinction between the two phases lies in the changes in their electronic structure near the band edges. At the same time, variations in the density of states close to the Fermi level affect both electrical conductivity and the generation of charge carriers. These effects are important for tuning the material's properties for solar energy conversion and microelectronic applications.

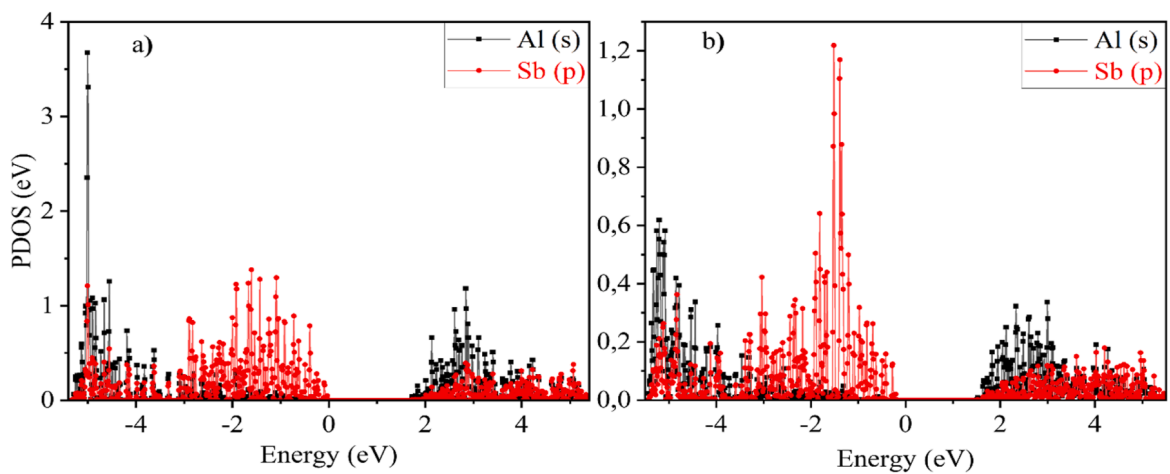

**Fig. 7.** Partial electronic density of states (PDOS) for the cubic (a) and hexagonal (b) phases of AlSb.

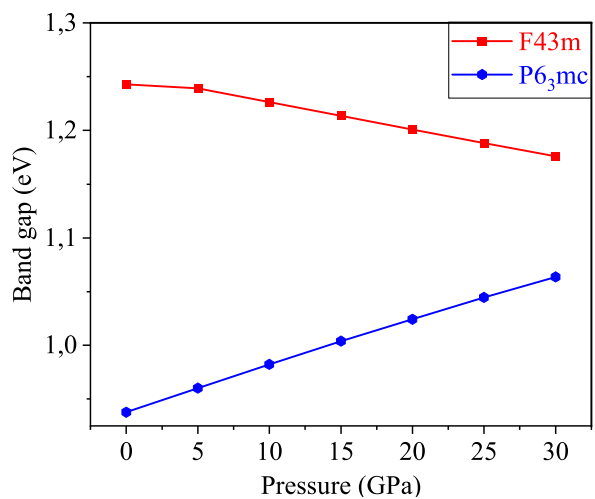

**Fig. 8.** Dependence of the band gap width of the cubic (red) and hexagonal (blue) phases of AlSb on external pressure, calculated in the range of 0–30 GPa.

To gain a deeper understanding of the contributions of individual atoms to the formation of the valence and conduction bands, the partial density of states (PDOS) was analyzed, as shown in Figs. 7a,b. The PDOS results reveal a clear contribution from the Sb p-orbitals and Al s-orbitals to the overall electronic structure. In both phases, the valence band is primarily composed of Sb p-orbitals, while the conduction band is mainly derived from Al s-orbitals.

In the cubic phase, a pronounced peak appears in the range from −4 to −5 eV, originating from the Al *s*-orbitals. This feature is absent in the hexagonal phase. The main contribution to the electronic density near the Fermi level comes from the Sb *p*-orbitals, which play a dominant role in defining the valence band. The PDOS spectra are in good agreement with the previously calculated band structures shown in Fig. 5.

### 3.3.1. Pressure-Dependent band gap evolution

In addition to the structural, phonon, thermodynamic, and electronic analyses, the variation of the electronic structure under external pressure was investigated to evaluate the evolution of the band gap and the mechanical and acoustic properties of AlSb. This approach provides insight into the stability of the material under mechanical stress. The calculated band gap values obtained using the GGA-PBE functional are summarized in Table S1 and reveal a clear and consistent trend. As shown in Fig. 8, in the cubic phase the band gap decreases with increasing external pressure within the 0–30 GPa range, while in the hexagonal phase it increases linearly with pressure. This opposite behavior arises from differences in the deformation response of the crystal lattices and the nature of the orbital hybridization between Al and Sb atoms.

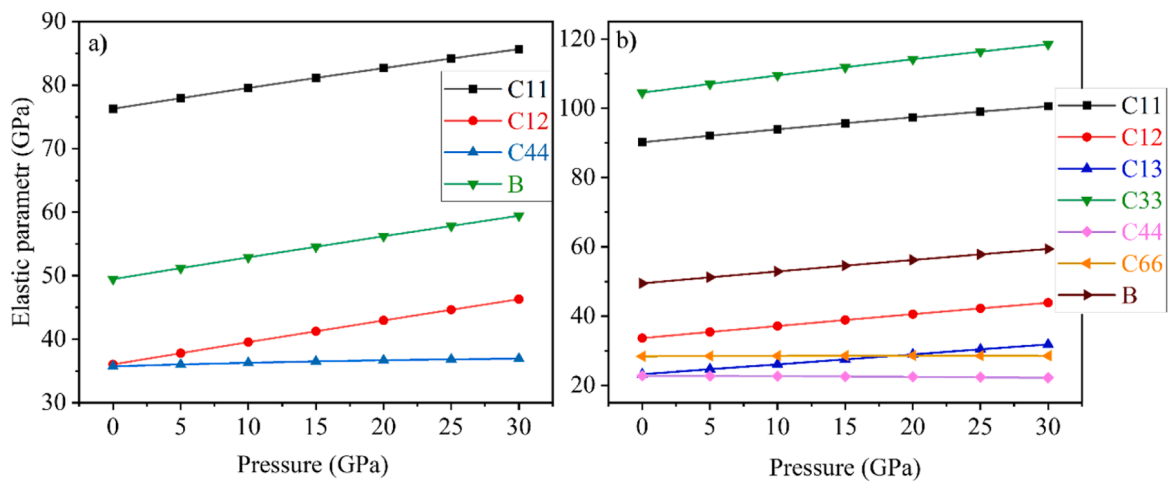

**Fig. 9.** Pressure dependence of the elastic constants $C_{ij}$ (GPa) and bulk modulus B for AlSb: (a) cubic phase and (b) hexagonal phase.

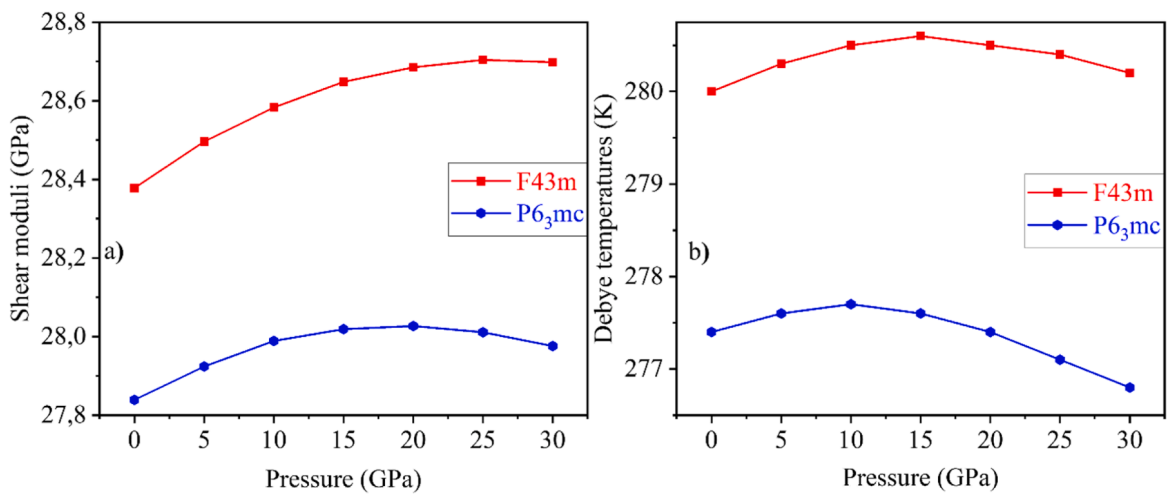

**Fig. 10.** Pressure dependence of (a) shear modulus (G) and (b) Debye temperature ($\Theta_D$) for AlSb.

In the cubic structure, compression enhances the overlap between Sb p-orbitals and Al s-orbitals, which lowers the conduction band minimum and reduces Eg. In contrast, in the hexagonal P6$_3$mc modification, changes in interatomic distances lead to a weakening of orbital overlap near the Γ point and to an increase in the energy separation between the valence and conduction bands. Overall, the opposite pressure-dependent trends of Eg reflect the different bonding characteristics and electronic-state symmetries of the two AlSb phases.

### 3.4. Mechanical properties and pressure effects

To understand how the pressure-induced evolution of electronic properties correlates with macroscopic stability, the mechanical characteristics of AlSb were calculated. Figs. 9a and 9b show the pressure-dependent elastic constants ($C_{11}$, $C_{12}$, $C_{44}$, etc.) and bulk modulus (B) for both phases in the range of 0–30 GPa. The complete set of elastic coefficients is listed in Table S2. The results indicate that both phases exhibit positive elastic constants ($C_{ij}$), satisfying the Born stability criteria for crystalline materials. This confirms that the AlSb system remains mechanically stable under the investigated pressure range, with no indication of shear instability or lattice softening. The increase in elastic moduli with pressure also suggests enhanced lattice stiffness and reduced compressibility, which is consistent with the observed pressure dependence of the band gap and phonon spectra.

As shown in Fig. 9, both phases demonstrate a consistent increase in the main elastic parameters with rising pressure. This effect is particularly evident for the longitudinal stiffness components $C_{11}$ and $C_{33}$, which reflect the growing rigidity of the lattice under compression. In contrast, the shear components $C_{44}$ and $C_{66}$ (for the hexagonal phase) remain nearly constant or slightly decrease, which can be attributed to the redistribution of internal stresses within the crystal structure.

Calculations were also carried out for the shear modulus (G), Debye temperature ($\Theta_D$), and acoustic velocities, including the transverse ($V_t$), longitudinal ($V_l$), and average sound velocities ($V_m$). The corresponding results are presented in Figs. 10a,b and 11a,b, and the complete numerical data are summarized in Table S3. These parameters are directly linked to the phonon properties and describe the stiffness of the lattice as well as its ability to sustain thermal vibrations. The increase in $\Theta_D$ and Vm with pressure reflects stronger interatomic bonding and reduced anharmonic effects, in full agreement with the trends observed in the phonon spectra.

According to the results, both phases exhibit a similar trend: the values of G and $\Theta_D$ increase with pressure up to approximately 15 GPa and then gradually decrease. Throughout the entire pressure range, the cubic phase maintains higher values of these parameters, which confirms its greater mechanical rigidity and thermal stability compared to the hexagonal modification. The observed behavior reflects the strengthening of interatomic interactions at moderate compression and their partial relaxation under higher external pressure, consistent with the general tendencies of covalent semiconductors.

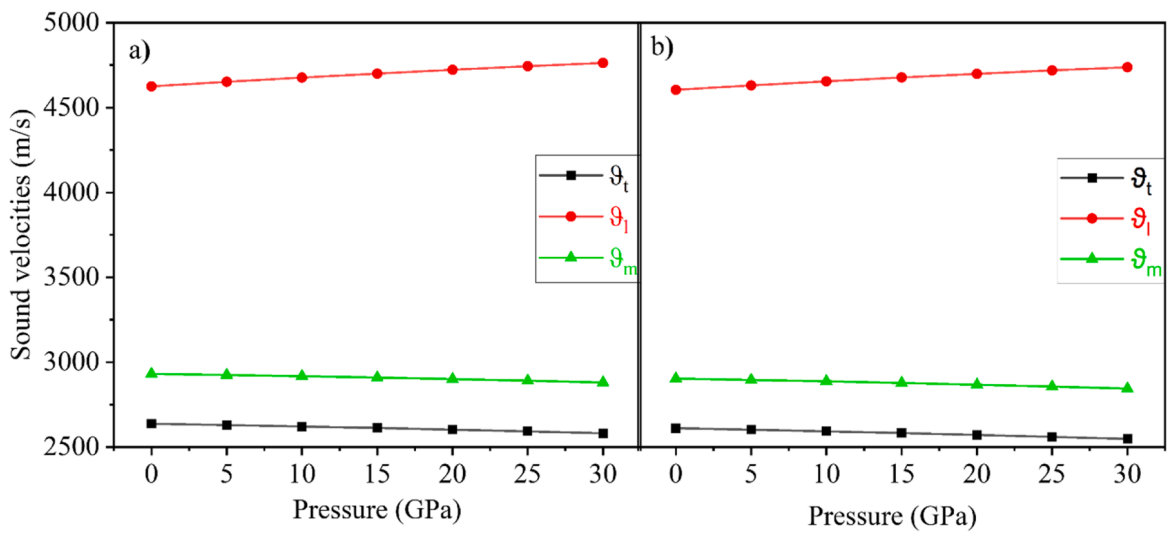

**Fig. 11.** Pressure dependence of longitudinal, transverse, and average sound velocities for (a) the cubic and (b) the hexagonal phases of AlSb.

**Table 5**
Pressure-dependent mechanical parameters of the cubic and hexagonal phases of AlSb obtained from GGA-PBE calculations.

| Mechanical properties AlSb | P (GPa) | B (GPa) | E (GPa) | v | M (GPa) | B/G |
|---|---|---|---|---|---|---|
| F43m | 0 | 49.457 | 71.466 | 0.259 | 87.295 | 1.743 |
| | 5 | 51.188 | 72.106 | 0.265 | 89.182 | 1.796 |
| | 10 | 52.893 | 72.661 | 0.271 | 91.004 | 1.850 |
| | 15 | 54.550 | 73.139 | 0.277 | 92.747 | 1.904 |
| | 20 | 56.206 | 73.544 | 0.282 | 94.453 | 1.959 |
| | 25 | 57.832 | 73.887 | 0.287 | 96.104 | 2.015 |
| | 30 | 59.433 | 74.157 | 0.292 | 97.697 | 2.071 |
| $P6_3mc$ | 0 | 49.452 | 70.321 | 0.263 | 86.571 | 1.776 |
| | 5 | 51.191 | 70.883 | 0.269 | 88.423 | 1.833 |
| | 10 | 52.867 | 71.371 | 0.275 | 90.185 | 1.889 |
| | 15 | 54.556 | 71.770 | 0.281 | 91.914 | 1.947 |
| | 20 | 56.197 | 72.095 | 0.286 | 93.566 | 2.005 |
| | 25 | 57.822 | 72.350 | 0.291 | 95.170 | 2.064 |
| | 30 | 59.396 | 72.539 | 0.296 | 96.698 | 2.123 |

The analysis of sound velocity as a function of pressure (Fig. 11, Table S3) provides deeper insight into the dynamic stability and acoustic stiffness of AlSb. Both phases show a clear tendency for $V_t$ and $V_m$ to decrease with increasing pressure from 0 to 30 GPa. This behavior is typical for dense crystalline materials where compression weakens shear-related phonon modes. In the cubic phase, the average sound velocity $V_m$ decreases from 2930 m/s to 2880 m/s, while the transverse velocity $V_t$ slightly drops from 2638 m/s to 2637 m/s, indicating a limited ability of the material to efficiently transmit shear vibrations.

In contrast, the longitudinal velocity $V_l$ shows a consistent increase, rising from 4625 m/s at 0 GPa to 4763 m/s at 30 GPa (Fig. 11a, Table S3). This trend can be attributed to enhanced bulk stiffness and the preservation of high density along longitudinal directions. The simultaneous decrease in $V_t$ and $V_m$ suggests a degree of anisotropy in the acoustic response under pressure. The cubic phase exhibits higher absolute values of all three velocity components compared with the hexagonal phase. At 0 GPa, $V_l$ reaches 4625 m/s, while $V_t$ and $V_m$ are 2637 m/s and 2931 m/s, respectively. Even at 30 GPa, these values remain relatively high (4763 m/s, 2582 m/s, and 2881 m/s), confirming the strong acoustic stiffness of the cubic modification and its potentially higher thermal conductivity. This feature is particularly advantageous for optoelectronic and thermoelectric applications, where efficient heat management is crucial for device reliability.

Based on the GGA-PBE calculations, additional mechanical properties, including the bulk modulus (B), Young's modulus (E), Poisson's ratio ($\nu$), longitudinal modulus (M), and Pugh's ratio (B/G), were evaluated and are summarized in Table 5. The results reveal that the cubic phase exhibits greater overall stiffness. A gradual increase in E, G, and B is observed with pressure, reflecting enhanced interatomic bonding and resistance to deformation. The B/G ratio exceeds 1.75 for both phases, which indicates ductile mechanical behavior and anisotropic bonding characteristics. The maximum value of B/G (2.123) is obtained for the hexagonal phase at 30 GPa, suggesting a partial softening of the lattice and an increase in plasticity under compression.

As shown in Table 5 and Figs. 2–4, the cubic phase exhibits higher values of the main mechanical and thermodynamic parameters across the entire pressure range. At 0 GPa, its bulk modulus reaches 49.5 GPa, and the Young's modulus is 71.5 GPa, both of which are greater than those of the hexagonal phase. With an increase in pressure to 30 GPa, the bulk modulus of the cubic phase rises to 59.4 GPa, indicating strong resistance to volumetric compression. Similarly, its shear modulus under ambient conditions is 28.4 GPa, and the transverse sound velocity equals 2637 m/s, both higher than those observed for the hexagonal structure.

The Debye temperature ($\Theta_D$) for the cubic phase reaches 280 K, compared to 277.4 K for the hexagonal phase. This difference reflects a stiffer phonon lattice and potentially higher thermal conductivity of the cubic structure. With increasing pressure up to 15 GPa, both the shear modulus (G) and Debye temperature ($\Theta_D$) gradually increase, followed by a slight decrease at higher pressures. Nevertheless, even at 30 GPa, the cubic phase maintains superior values of these parameters, confirming its high mechanical and thermal robustness. Fig. 12 presents the pressure dependence of the lattice parameters (a, b, c) and unit cell volume for both AlSb phases. As the pressure increases, all lattice parameters decrease nearly linearly, consistent with the compressibility behavior typical of crystalline solids. In the cubic phase, compression leads to a more pronounced reduction in volume compared with the hexagonal phase, indicating higher compressibility of the cubic structure. In contrast, the hexagonal modification exhibits nearly uniform contraction along the a, b, and c axes, reflecting moderate anisotropy and stability against structural distortions under compression.

Overall, both AlSb phases demonstrate elastic stability within the investigated pressure range of 0–30 GPa. However, the hexagonal structure shows slightly lower deformation rates and potentially greater mechanical reliability under sustained pressure, while the cubic phase remains the stiffer and thermally more conductive polymorph.

The results shown in Fig. 12 illustrate stable and physically consistent behavior of both AlSb phases under external pressure. The in-plane and out-of-plane lattice parameters, as well as the unit cell volume, decrease systematically with increasing pressure from 0 to 30 GPa. This trend reflects the elastic compressibility of the crystal lattice and confirms the absence of any pressure-induced phase transitions within the investigated range.

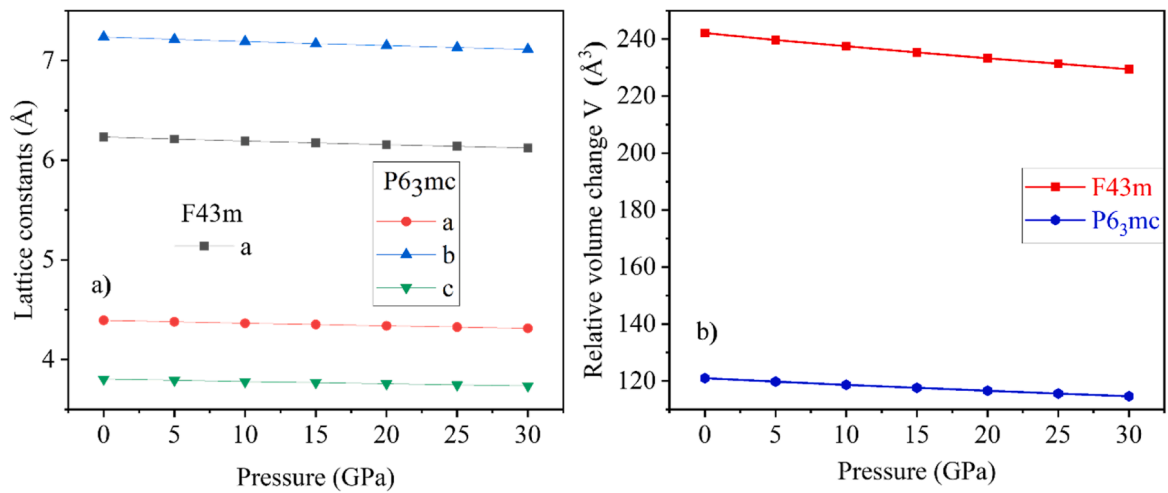

**Fig. 12.** Pressure dependence of the lattice parameters and relative unit cell volume of AlSb: (a) lattice constants, (b) relative volume variation.

*3.5. Optical properties*

To further evaluate the optical response of AlSb, the real ($\varepsilon_1$) and imaginary ($\varepsilon_2$) parts of the dielectric function were calculated for both the cubic F-43 m and hexagonal P6$_3$mc phases (Figures S3 and S4). Optical phenomena including absorption coefficient ($\alpha(\omega)$), energy loss function ($L(\omega)$), extinction coefficient ($k(\omega)$), reflectivity ($R(\omega)$), and refractive index ($n(\omega)$) are present in all types of materials and can be studied at both microscopic and macroscopic levels. In bulk materials, the complex permittivity is closely related to the band structure at the microscopic or quantum mechanical level [60–62]. Optical properties such as the refractive index and absorption coefficient indicate the response of the material when photons hit them. The real part of the permittivity, which is considered an inherent property of any material, indicates the stored energy that can be released at zero energy or frequency limit. Using the Ab initio calculation algorithms implemented in the VASP code, the complex imaginary permittivity can be obtained in the PAW method by summation over conduction bands, where the expression obtained in [63] for determining $\varepsilon_2(\omega)$ has the following form:

$$\varepsilon_2^{\alpha\beta}(\omega) = \frac{4\Pi^2 e^2}{\Omega} \frac{1}{q^2} \lim_{q \to 0} \sum_{c,\vartheta,k} 2\omega_k \delta(\epsilon_{ck} - \epsilon_{\vartheta k} - \omega) \times \langle u_{ck+e_\alpha q} | u_{ck} \rangle \langle u_{ck+e_\alpha q} | u_{\vartheta k} \rangle^* \quad (1)$$

In the above equation the transitions are made from occupied to unoccupied states within the first Brillouin zone, the wave vectors are fixed k. Real ($\varepsilon_1$) and Imaginary ($\varepsilon_2$) parts of analytical dielectric function are connected by the Kramers - Kronig relationas [64],

$$\varepsilon_1^{\alpha\beta}(\omega) = 1 + \frac{2}{\Pi} P \int_0^\infty \lim_{q \to 0} \frac{\varepsilon_2^{\alpha\beta}(\omega')\omega'}{\omega'^2 - \omega^2 + in} d\omega' \quad (2)$$

Given Eq. (1–2), the optical conductivity spectrum ($\sigma$), energy loss spectrum (L), refractive index (n), reflection coefficient (R), absorption coefficient ($\alpha$), and extinction coefficient (k) may be calculated as follows:

$$R(\omega) = \left| \frac{\sqrt{\varepsilon(\omega)} - 1}{\sqrt{\varepsilon(\omega)} + 1} \right|^2 \quad (3)$$

where $\varepsilon(\omega) = \varepsilon_1(\omega) + i\varepsilon_2(\omega)$.

$$L(\omega) = \frac{\varepsilon_2(\omega)}{\varepsilon_1^2(\omega) + \varepsilon_2^2(\omega)} \quad (4)$$

$$n(\omega) = \left| \frac{\sqrt{\varepsilon_1^2 + \varepsilon_2^2} + \varepsilon_1}{2} \right|^{\frac{1}{2}} \quad (5)$$

$$k(\omega) = \left| \frac{\sqrt{\varepsilon_1^2 + \varepsilon_2^2} - \varepsilon_1}{2} \right|^{\frac{1}{2}} \quad (6)$$

$$\alpha(\omega) = \frac{2\omega k(\omega)}{c} = \frac{\varepsilon_2(\omega)\omega}{n(\omega)c} \quad (7)$$

$$\sigma(\omega) = \sigma_1(\omega) + i\sigma_2(\omega) = -i\frac{\omega}{4\pi}[\varepsilon(\omega) - 1] = \frac{\varepsilon_2(\omega)\omega}{4\pi} + i\frac{1 - \varepsilon_1(\omega)\omega}{4\pi} \quad (8)$$

From these quantities, key optical parameters were derived, including the absorption coefficient ($\alpha(\omega)$), energy loss function ($L(\omega)$), extinction coefficient ($k(\omega)$), reflectivity ($R(\omega)$), and refractive index ($n(\omega)$). Fig. 13 (a-c) presents the computed spectra of the absorption coefficient $\alpha(\omega)$, the energy loss function $L(\omega)$, and the extinction coefficient $k(\omega)$ for both AlSb phases. The results show typical semiconductor-like behavior, where the absorption onset occurs in the range of 1.6–1.7 eV, consistent with the calculated band gap. In the low-energy region, the sharp rise in $\alpha(\omega)$ indicates direct interband transitions, while multiple pronounced peaks appear between 4–8 eV, corresponding to electronic transitions from the Sb p-states in the valence band to the Al s/p-states in the conduction band. The maximum absorption coefficient $10^6$ cm$^{-1}$ is comparable to or higher than that of GaSb, InSb, and AlP thin films reported in recent works [65–72], indicating that AlSb can efficiently absorb UV–visible photons.

The stronger maxima observed for the hexagonal phase indicate a higher density of accessible states near the Fermi level, which enhances the probability of optical transitions and accounts for its stronger high-energy response. The energy loss function (Fig. 13b) shows a dominant plasmon peak near 13 eV for both phases, corresponding to collective oscillations of the electron gas and revealing similar plasma energies for the two structures. The extinction coefficient (Fig. 13c) follows the general trend of the absorption spectrum, peaking in the 4–6 eV range and gradually decreasing at higher photon energies, reflecting the dispersive nature of the optical response. The absorption coefficients of both phases reach magnitudes on the order of $10^6$ cm$^{-1}$, which is typical for highly efficient light-harvesting materials and makes AlSb a strong candidate for thin-film photovoltaic applications. In the photon energy range of 2–12 eV, both phases display multiple resonance peaks associated with interband transitions. Overall, the AlSb phases exhibit high transparency in the visible region and strong absorption in the ultraviolet range, making them promising for photonics and optoelectronic device applications such as photodetectors, LEDs, and solar absorbers. It should be noted that the present optical calculations do not explicitly include excitonic effects, optical anisotropy, or local-field effects. These factors may slightly modify the absorption edge, especially in the ultraviolet region. However, previous studies on III–V semiconductors show that such effects mainly lead to minor spectral shifts and do not change the general absorption trends.

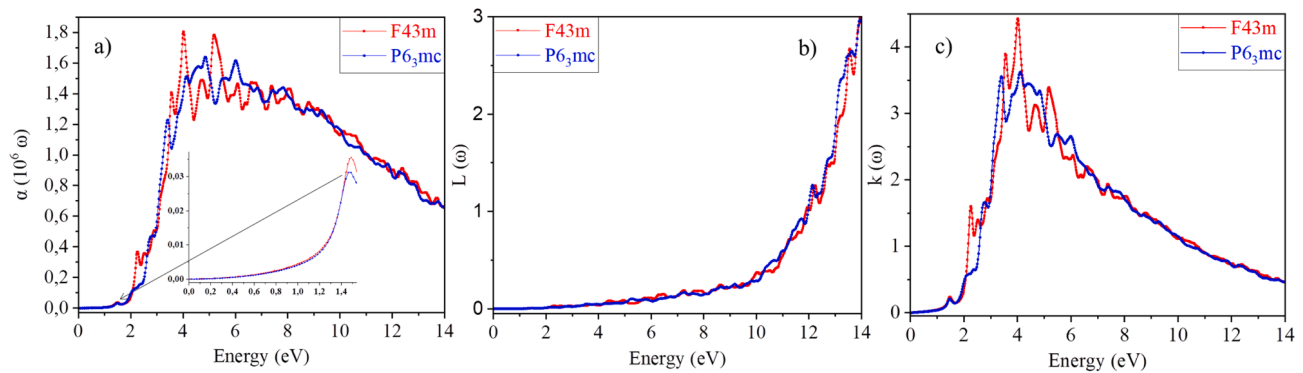

**Fig. 13.** Calculated spectra of (a) absorption coefficient $\alpha(\omega)$, (b) energy loss function $L(\omega)$, and (c) extinction coefficient $k(\omega)$ for cubic (red) and hexagonal (blue) phases of AlSb.

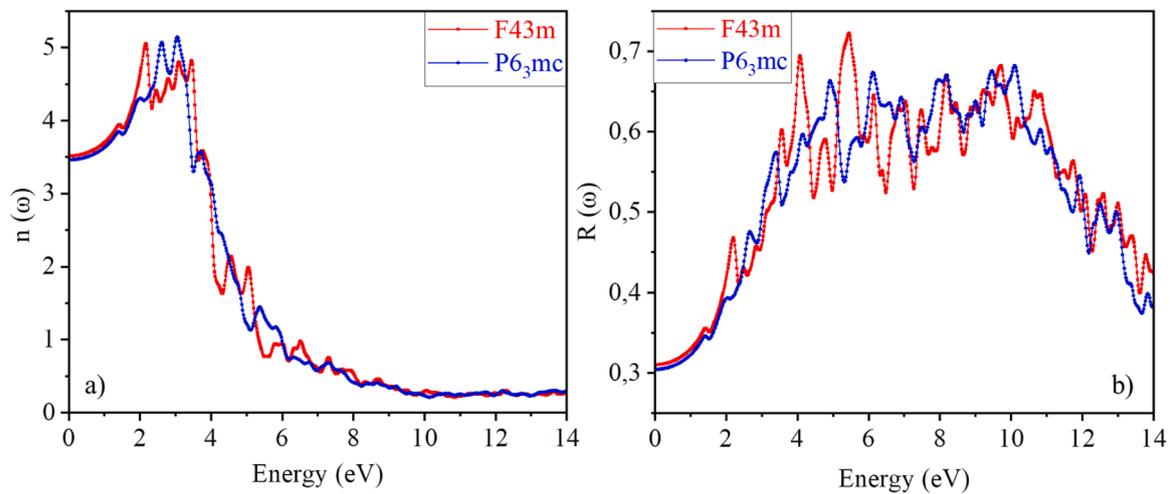

**Fig. 14.** Calculated spectra of (a) refractive index $n(\omega)$ and (b) reflectivity $R(\omega)$ for cubic (red) and hexagonal (blue) phases of AlSb.

**Table 6**
Static optical parameters of cubic and hexagonal phases of AlSb, including the refractive index n(0), real part of the dielectric constant $\varepsilon_1(0)$, and reflectivity R(0).

| AlSb | $\varepsilon_1(0)$ | n(0) | R(0) |
| --- | --- | --- | --- |
| F43m | 12.344 | 3.51 | 0.310 |
| P6$_3$mc | 11.994 | 3.46 | 0.305 |

Fig. 14 shows the spectra of the refractive index $n(\omega)$ and reflectivity $R(\omega)$ for the F-43 m and P6$_3$mc phases of AlSb, with their static values summarized in Table 6. In the low-energy region, both modifications display the typical semiconductor behavior characterized by a sharp rise in $n(\omega)$ as the photon energy approaches the optical absorption threshold. The maximum refractive index for the cubic phase reaches approximately $n_{max} \approx 5.2$ at about 2.1 eV, while for the hexagonal phase, it is slightly higher ($n_{max} \approx 5.4$ at 2.0 eV).

These values are comparable to those reported for other III-V semiconductors such as InSb and GaSb, reflecting the high polarizability and strong covalent bonding nature of the Al-Sb system. The high refractive index suggests strong light-matter interaction, which is beneficial for enhancing optical confinement in nanoscale devices. Additionally, the reflectivity spectra exhibit moderate values (up to 60 %), indicating efficient coupling of incident light into the material without excessive surface reflection, a desirable feature for photovoltaic and optoelectronic applications where both absorption and internal photon management are crucial

The reflectivity coefficient $R(\omega)$ (Fig. 14b) exhibits a similar energy-dependent trend, with a maximum of $R_{max} \approx 0.73$ for the cubic F-43 m phase and $\approx 0.75$ for the hexagonal P6$_3$mc phase in the 2–3 eV range, followed by a gradual decline with increasing photon energy. The high reflectivity values observed in the visible and near-UV regions highlight the potential of AlSb as a material for optoelectronic coatings and light-reflective structures. Such behavior indicates efficient reflection of incident radiation at moderate photon energies while maintaining adequate transparency at higher energies, making AlSb suitable for multi-layer optical filters, solar concentrators, and infrared detector windows. The combination of a large refractive index and strong reflectivity confirms its promise for integration into photonic and energy-conversion devices, where precise control of light propagation and reflection is essential.

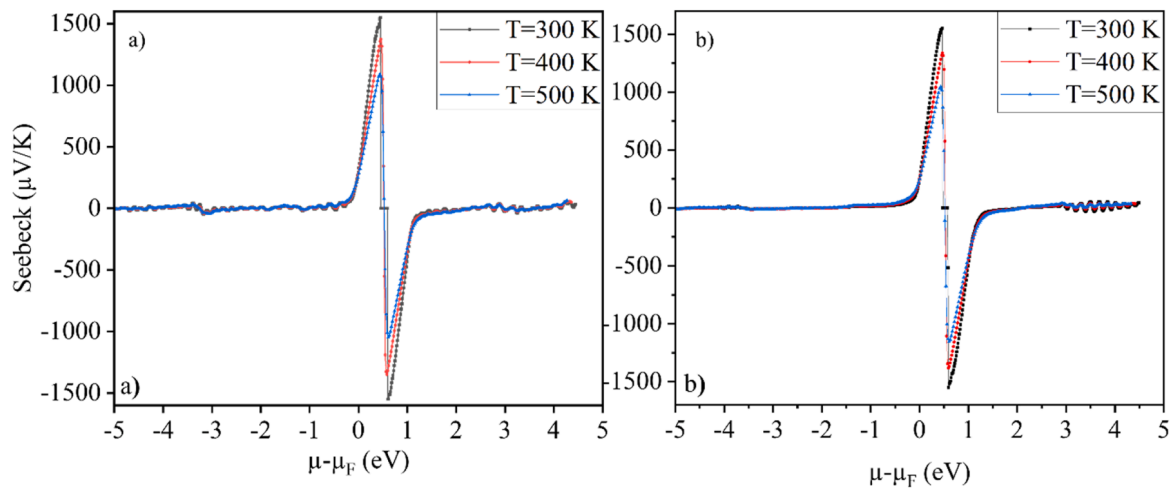

**Fig. 15.** Seebeck coefficient for (a) the cubic and (b) the hexagonal phases of AlSb at 300 K, 400 K, and 500 K.

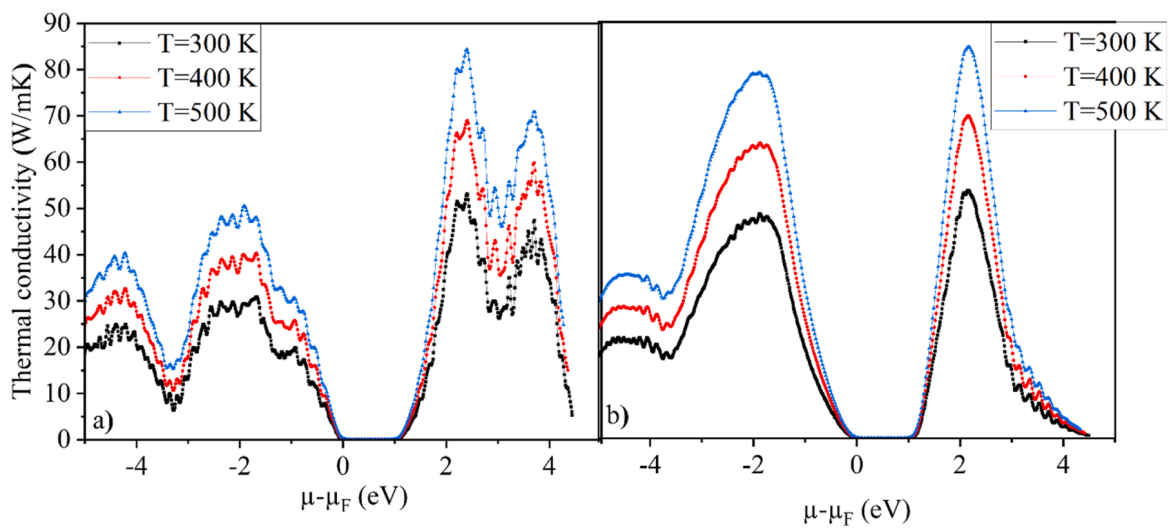

**Fig. 16.** Thermal conductivity of (a) cubic and (b) hexagonal AlSb phases at 300 K, 400 K, and 500 K.

From the calculated static optical constants $n(0)$, $\varepsilon_1(0)$, and $R(0)$, it can be observed that the hexagonal modification exhibits a slightly stronger optical response and higher dielectric polarizability than the cubic phase. These findings are consistent with the band gap values obtained using the mBJ functional and listed in Table 4. With increasing $E_g$, the values of $\varepsilon$, $n$, and $R$ gradually decrease. This trend agrees with the general Penn model, which relates a larger band gap to lower dielectric polarizability. The consistency between the electronic and optical results confirms the reliability of the computational approach and the accuracy of the calculated data. The observed correlation between optical parameters and electronic structure highlights the intrinsic connection between chemical composition, density of states, band gap width, and optical response of the material.

### 3.6. Thermoelectric properties

The Seebeck effect occurs when two different conductors or semiconductors are connected in series under a temperature gradient between their junctions [73–75]. This gradient drives the diffusion of electrons from the hot region toward the cold region, generating a thermoelectromotive force typically measured in microvolts per kelvin. The magnitude of this potential depends on the material type and the transport properties of charge carriers. The quality of a thermoelectric device is determined primarily by the Seebeck coefficient, where a larger coefficient indicates a higher efficiency in converting thermal energy into electricity.

In this study, the thermoelectric properties of AlSb were analyzed to understand how crystal structure and temperature influence the Seebeck coefficient (S), electrical conductivity ($\sigma$), electronic thermal conductivity ($\kappa_e$), and power factor (PF). These parameters were calculated using semiclassical Boltzmann transport theory as implemented in BoltzTraP:

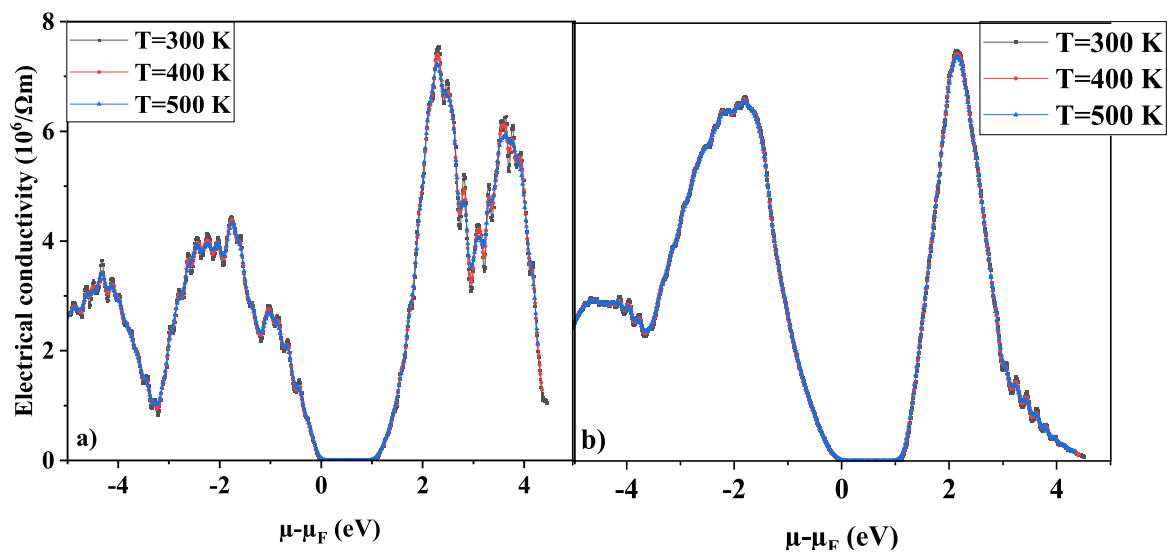

**Fig. 17.** Electrical conductivity of (a) cubic and (b) hexagonal AlSb phases at 300 K, 400 K, and 500 K.

$$S = \left(\frac{1}{eT}\right) \cdot \left(\frac{K_1}{K_0}\right),$$

$$\sigma = e^2 K_0,$$

$$k_e = \left(\frac{1}{T}\right) \cdot \left(\frac{K_2 - K_1^2}{K_0}\right),$$

$$PF = S^2 \sigma$$

where e is the elementary charge, T is the absolute temperature, and $K_0$, $K_1$, and $K_2$ are the transport integrals derived from the electronic band structure [61].

The Seebeck coefficient is highly sensitive to the electronic states near the Fermi level, and its magnitude and sign depend on the carrier type, effective mass, curvature of the band edges, and the position of the chemical potential. Evaluating S as a function of the Fermi level provides direct insight into intrinsic thermoelectric behavior and optimal doping conditions [76–79]. The dependence of S on chemical potential at 300, 400, and 500 K for both phases is shown in Figs. 15a and 15b The interaction between electronic excitations and lattice vibrations observed in the optical spectra also plays a decisive role in the thermoelectric performance of AlSb. Materials with high dielectric constants and strong light–matter coupling often demonstrate enhanced carrier mobility and tunable electron–phonon scattering, improving both the Seebeck coefficient and electrical conductivity. Therefore, analysis of optoelectronic properties provides key insight into microscopic mechanisms governing charge and heat transport in both phases.

Both polymorphs exhibit strong thermopower, with the p-type cubic and hexagonal structures reaching approximately 1550 and 1545 μV/K at 300 K, respectively. These values exceed the typical temperature-dependent Seebeck coefficients of bulk III-V semiconductors (300–600 μV/K) but are consistent with previous first-principles studies where S is computed as a function of the chemical potential rather than temperature [80–87], which often yields upper-limit theoretical values compared with classical temperature-dependent estimates [86,87]. In our calculations, the transport coefficients were obtained using the BoltzTraP2 code within the constant relaxation time approximation (CRTA), where the relaxation time τ is assumed to be independent of energy and temperature. This approximation, combined with the chemical-potential-based formulation of S, can enhance Seebeck magnitudes, particularly in materials with narrow carrier pockets or pronounced band-edge anisotropy. Therefore, the Seebeck coefficients exceeding 1500 μV/K reported here should be regarded as intrinsic theoretical upper bounds rather than directly measurable experimental values. Nevertheless, the temperature-driven behavior of S and the favorable balance between thermopower and electrical conductivity clearly confirm the intrinsic thermoelectric potential of both AlSb phases. Thus, the reported Seebeck coefficients above 1500 μV/K should be viewed as intrinsic theoretical limits arising from the CRTA-based Boltzmann transport formulation rather than experimentally attainable values. As temperature rises, the coefficient increases due to enhanced carrier excitation, as shown in Figure S5. For the cubic phase, the Seebeck coefficient grows monotonically, while in the hexagonal phase it remains nearly zero between 150–250 K and begins to rise only above 250 K. This indicates that thermoelectric activity in the hexagonal structure emerges at higher temperatures compared to the cubic one.

In semiconductors, thermal conductivity is primarily governed by phonon transport, whereas in metals it is dominated by free electrons. For efficient thermoelectric materials, low thermal conductivity is essential to maintain a strong temperature gradient. Fig. 16 (a, b) shows the calculated thermal conductivity of both AlSb phases at 300 K, 400 K, and 500 K as a function of chemical potential. The reduction of thermal conductivity in both AlSb phases, especially above 600 K, is closely related to enhanced lattice anharmonicity and phonon–phonon scattering. At elevated temperatures, the increased vibrational amplitude of atoms strengthens three-phonon and higher-order scattering channels, which suppresses heat transport by shortening the phonon mean free path. This effect is more pronounced in the hexagonal P6$_3$mc phase due to its lower lattice stiffness and higher anisotropy, which reduce the group velocity of acoustic phonons. As a result, κ decreases more strongly with temperature, enabling higher thermoelectric efficiency by increasing the S$^2$σ/κ ratio. These observations are consistent with recent studies on anharmonic transport in AlSb and related III–V semiconductors.

Both phases exhibit nearly identical behavior. At room temperature, in the range 0.46–0.59 eV, the thermal conductivity for both structures is close to zero, corresponding to the optimal chemical potential range for maximizing thermoelectric efficiency. As the chemical potential moves away from this region, the thermal conductivity begins to increase, reducing the thermoelectric performance. To gain a deeper understanding of thermal transport, the temperature dependence of thermal conductivity was analyzed in the range from 150 to 1000 K, as shown in Figure S6. In the cubic phase, the thermal conductivity remains nearly zero up to about 400 K, after which it gradually increases. In the hexagonal phase, it stays close to zero up to approximately 600 K before showing a noticeable rise. At 300 K, the thermal conductivity of both phases is effectively zero, indicating very weak phonon-mediated heat transport at low temperatures.

When free electrons absorb heat, they gain kinetic energy and migrate toward the colder region of the material, generating an electric current. For a thermoelectric device to operate efficiently, the material must exhibit high electrical conductivity in order to minimize Joule heating losses. Figs. 17a, b presents the calculated electrical conductivity of cubic and hexagonal AlSb phases at 300 K, 400 K, and 500 K. Both materials show similar temperature-dependent trends.

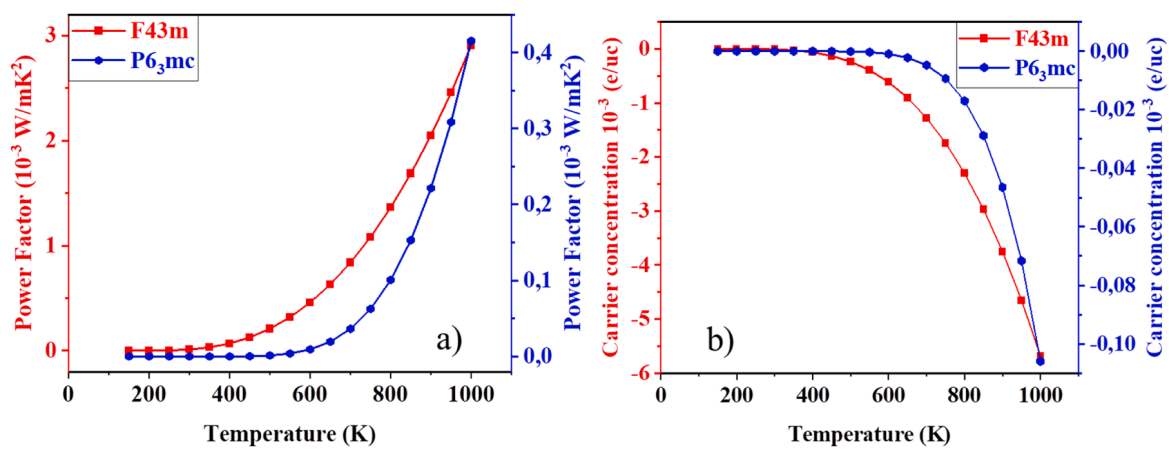

**Fig. 18.** (a) Temperature dependence of the power factor and (b) charge carrier concentration for the cubic (red) and hexagonal (blue) phases of AlSb.

The plots reveal that the key chemical potential range for optimal conductivity in both p-type and n-type configurations lies between 0.46 and 0.59 eV. Beyond this range, electrical conductivity decreases, which defines the region where AlSb exhibits favorable thermoelectric performance. Within the specified window, both phases display a stable conductive behavior, confirming that AlSb maintains efficient charge transport properties across a broad range of chemical potentials.

The temperature dependence of electrical conductivity in the 150–1000 K range is shown in Figure S7. According to the results, the cubic phase of AlSb begins to exhibit noticeable conductivity at around 450 K, while the hexagonal modification becomes conductive near 650 K. As the temperature increases, the electrical conductivity of both phases grows exponentially due to thermal activation of charge carriers and a reduction in the band gap width. At 800 K, the electrical conductivity of the cubic phase reaches approximately $1.0 \times 10^4$ (S·m)$^{-1}$, whereas that of the hexagonal phase is about $7.5 \times 10^3$ (S·m)$^{-1}$, indicating higher carrier mobility and concentration in the cubic structure. The enhanced conductivity of the F-43 m phase correlates with its lower effective carrier mass and higher crystal symmetry, which facilitate more efficient charge transport.

The power factor (PF) reflects the ability of a material to convert thermal energy into electrical energy efficiently and depends on the balance between the Seebeck coefficient and electrical conductivity. As shown in Fig. 18a, both AlSb phases behave as typical semiconductors at low temperatures, where the concentration of thermally activated carriers is low and PF values are nearly zero. Starting at approximately 300 K for the cubic phase and 450 K for the hexagonal modification, a gradual increase in the power factor is observed, driven by carrier activation and improved electrical transport. At temperatures above 800 K, the hexagonal structure exhibits comparable or slightly higher PF values, which can be attributed to its lower thermal conductivity and enhanced phonon-electron scattering that suppresses carrier backflow. This temperature-dependent behavior confirms that AlSb is a promising candidate for high-temperature thermoelectric applications, where large thermal gradients are present. The observed trends are consistent with current research showing that for improving thermoelectric device efficiency above 600 K, it is essential to minimize thermal conductivity while maintaining a high $S^2\sigma / \kappa$ ratio.

Fig. 18b shows the temperature dependence of the charge carrier concentration. At low temperatures, the carrier density in both phases is close to zero, confirming the semiconducting nature of AlSb. As temperature increases, the carrier concentration rises steadily, following the expected thermal excitation trend typical for narrow-gap semiconductors, which further supports its potential use in high-efficiency thermoelectric converters and infrared energy-harvesting devices.

As temperature increases, the carrier concentration grows exponentially, particularly above 600 K, which is consistent with the mechanism of thermal generation of electron-hole pairs. The steeper increase observed for the cubic phase reflects its smaller band gap and higher density of states near the conduction band edge, resulting in stronger carrier excitation.

The combined analysis of Table S4, Figs. 15–18, and Figures S5-S7 demonstrates that the hexagonal P6$_3$mc phase of AlSb exhibits potentially higher thermoelectric efficiency at elevated temperatures. This improvement arises from the combination of high thermal stability, moderate electrical conductivity, and reduced lattice thermal conductivity, factors that are central to the optimization strategies of modern thermoelectric materials such as Bi$_2$Te$_3$, SnSe, and transition-metal oxynitrides. These results indicate that structural tuning and phonon engineering in AlSb-based systems can further enhance their high-temperature thermoelectric performance and broaden their applicability in energy conversion technologies. Finally, the correlation between crystal symmetry, phonon dispersion, and charge transport in AlSb becomes evident when comparing the two phases. The higher symmetry of the cubic F-43 m structure results in more dispersive acoustic phonon branches and lower effective masses, which enhance carrier mobility and electrical conductivity. In contrast, the lower symmetry and anisotropic bonding environment of the hexagonal P6$_3$mc phase lead to flatter phonon modes, reduced group velocities, and stronger phonon scattering, which suppress thermal conductivity and improve thermoelectric efficiency at high temperatures. This interplay between structural symmetry, lattice dynamics, and electronic transport is a key factor governing the distinct thermoelectric responses of the cubic and hexagonal phases of AlSb.

## 4. Conclusion

A detailed theoretical analysis of AlSb in its cubic and hexagonal phases has been carried out to elucidate the interplay between structural, thermodynamic, electronic, optical, and thermoelectric characteristics that define its multifunctional potential. Both modifications were confirmed to be dynamically and mechanically stable, with the cubic phase exhibiting slightly higher thermodynamic stability, as supported by its lower formation energy and stronger interatomic bonding network. Phonon dispersion, elastic constants, and Debye temperature analyses indicate robust lattice integrity and favorable thermal behavior up to 1000 K. Electronic structure calculations revealed that both phases possess a direct band gap in the optimal range for infrared and solar-energy conversion technologies, while the pressure-dependent band-gap evolution underscores the possibility of band-structure engineering through strain or epitaxial control.

Optical modeling revealed strong absorption and high refractive indices comparable to established III-V semiconductors, suggesting AlSb's potential in infrared photodetectors, light-emitting diodes, and high-frequency optoelectronic devices. Thermoelectric simulations further showed that the hexagonal phase, owing to its reduced lattice stiffness and lower thermal conductivity, may outperform the cubic phase at elevated temperatures, highlighting AlSb as a viable candidate for energy-harvesting and waste-heat recovery systems.

Overall, this work bridges microscopic bonding characteristics and macroscopic functionalities, offering a comprehensive picture of AlSb as a dual-purpose semiconductor, combining optical transparency, electronic tunability, and thermoelectric efficiency. The insights presented herein provide a theoretical framework for future experimental efforts and guide the rational design of III-V-based materials for integrated optoelectronic and energy-conversion devices.

**Funding**

This work was supported by the Interstate Fund for Humanitarian Cooperation of the CIS Member States through the scientific project funded by the International Nanotechnology Innovation Center of the CIS (grant no. 25–113), and by the International Science and Technology Center (ISTC) (grant no. TJ-2726).

**CRediT authorship contribution statement**

**Iskandar Raufov:** Validation, Resources, Methodology, Investigation, Formal analysis, Data curation. **Dilshod Nematov:** Writing – review & editing, Writing – original draft, Supervision, Project administration, Investigation, Formal analysis. **Saidjafar Murodzoda:** Methodology, Data curation. **Sakhidod Sattorzoda:** Writing – original draft, Visualization, Software, Resources, Investigation, Formal analysis. **Anushervon Ashurov:** Writing – original draft, Funding acquisition, Conceptualization.

**Declaration of competing interest**

The authors declare that they have no known competing financial interests or personal relationships that could have appeared to influence the work reported in this paper.

**Supplementary materials**

Supplementary material associated with this article can be found, in the online version, at doi:10.1016/j.ctta.2025.100255.

**Data availability**

No data was used for the research described in the article.